\documentclass[aps,twocolumn,showpacs]{revtex4}
\usepackage{epsfig}
\usepackage{graphicx}
\usepackage{textcomp}

\usepackage{epstopdf}
\usepackage{amsmath, amsthm, amssymb}
\usepackage{bm}

\usepackage[colorlinks=true, urlcolor=blue, anchorcolor=blue, citecolor=blue,filecolor=blue,linkcolor=blue,menucolor=blue,pagecolor=blue]{hyperref}

\newcommand{\vect}[1]{\bm {#1}} % for vectors
\newcommand{\mat}[1]{\mathbb {#1}} % for matrices

\begin{document}
\title{Extinction of oscillating populations}

\author{Naftali R. Smith$^{1}$ and Baruch Meerson$^{1}$}
\affiliation{$^{1}$Racah Institute of Physics, Hebrew University of Jerusalem, Jerusalem 91904, Israel}

\begin{abstract}

Established populations often exhibit oscillations in their sizes. If a population is isolated, intrinsic stochasticity of elemental processes can ultimately bring it to extinction. Here we study extinction of oscillating populations in a stochastic version of the Rosenzweig-MacArthur predator-prey model. To this end we extend a WKB approximation (after Wentzel, Kramers and  Brillouin) of solving the master equation to the case of extinction from a limit cycle in the space of population sizes. We evaluate the extinction rates and find the most probable paths to extinction by applying Floquet theory to the dynamics of an effective WKB Hamiltonian. We show that the entropic barriers to extinction change in a non-analytic way as the system   passes through the Hopf bifurcation. We also study the subleading pre-exponential factors of the WKB approximation.

\end{abstract}

%87.18.Tt	Noise in biological systems
%87.23.Cc	Population dynamics and ecological pattern formation
%02.50.Ga	Markov processes
%05.40.Ca	Noise
\maketitle

\section{Introduction}
\label{sec:intro}

Populations of individuals (of molecules, bacteria, animals or even humans) can often be viewed as stochastic. The intrinsic (demographic) noise in the elemental processes, governing these systems, profoundly affects their dynamics. A dramatic example is extinction of a long-lived isolated population resulting from a rare sequence of events when deaths prevail over births.
Stochastic population dynamics in general, and population extinction in particular, have always been a part of population biology \cite{OvMe2010}. More recently they have attracted attention from statistical physicists, who
view stochastic populations as a many-body system far from thermal equilibrium.

The intrinsic-noise-driven extinction of single populations is by now well understood, see Refs. \citep{OvMe2010,Assaf2010metastable} and references therein. Extinction of one or more populations in long-lived multi-population systems  has been also extensively studied, assuming that, prior to extinction, the populations reside in the vicinity of an attracting fixed point in the space of population sizes  \citep{Dykman2008,Kamenev2008,KD2009,KMS2010,KDM2010,Lohmar2011,GM2012,GMR2013}. Many co-existing populations, however, exhibit persistent oscillations in their sizes \citep{Elton1942, Butler1953, Odum2004, Murray2008}. At the level of deterministic theory, these oscillations are usually described by a stable limit cycle in the space of population sizes. A well-known deterministic model that shows this feature -- a variation of the celebrated Lotka-Volterra model \citep{Lotka1920, Volterra1927} -- is due to Rosenzweig and MacArthur \citep{RM1963}. Qualitatively similar models are used  in epidemiology -- for a description of the oscillatory dynamics of susceptible and infected populations during an epidemic  \citep{bartlett1960stochastic_t, Anderson1991_t, Andersson2000_t} and the oscillatory dynamics of tumor growth \citep{Kirschner1998, d'Onofrio2009}. Similar models describe oscillatory chemical reactions  \citep{Schlogl1972, VanKampen2007}.

In this work we study extinction of oscillating populations driven by intrinsic noise. We evaluate the extinction rates and most likely routes to extinction in a stochastic version of the Rosenzweig-MacArthur (RMA)  model  that we suggest. We extend a WKB theory (after Wentzel, Kramers and  Brillouin), previously
developed for multiple populations residing in the vicinity of an attracting fixed point \citep{Dykman2008,Kamenev2008,KD2009,KMS2010,KDM2010,Lohmar2011,GM2012,GMR2013,Dykman1994,RFRSM2005}, to populations residing in the vicinity of a stable limit cycle.  We show that the most likely routes to extinction in such systems are described by a new type of instantons -- special  phase trajectories of the underlying effective classical mechanics.
In its leading order, the WKB theory yields the mean time to extinction (MTE) with an exponential accuracy, that is up to a sub-leading prefactor. As we show here, our WKB results agree with numerical solutions of the master equation, and with direct Monte-Carlo simulations for this model, up to a prefactor, which is a subleading correction to the WKB result.
We find that the entropic barrier to extinction behaves in a non-analytic way at the Hopf bifurcation describing the birth of limit cycle. Furthermore, we evaluate the subleading WKB prefactor numerically and find that it too changes its behavior at the Hopf bifurcation. We suggest theoretical arguments to explain these features.
The results of this work can be extended to a whole class of models of isolated multiple populations which exhibit, at the deterministic level, a stable limit cycle.

Here is how the remainder of the paper is structured. In Sec. \ref{sec:model} we briefly recap the main properties of  the deterministic RMA model and focus on the parameter region where the system exhibits a stable limit cycle in the space of population sizes. Section \ref{sec:stochastic} deals with the stochastic version of the RMA model that we suggest. In particular, subsection \ref{sec:stochastic:two}  discusses the two routes to
extinction that this model exhibits. In subsection \ref{sec:stochastic:master}
we present the master equation for the stochastic version of the RMA model and discuss its long-time properties. In subsection \ref{sec:stochastic:WKB}
we develop a WKB theory of population extinction in this model. We summarize our results in Sec. \ref{sec:discussion}.

\section{Rosenzweig-MacArthur Model: Deterministic Dynamics and Limit Cycle}
\label{sec:model}

We denote the population sizes of the predators and prey by $F$ (foxes) and $R$ (rabbits), respectively, and assume that the population densities are homogeneous in space.
The elemental processes and rates of the Rosenzweig-MacArthur (RMA) model \cite{RM1963} are presented in Table \ref{table:reactions}. The rabbits reproduce at rate $a$, and die, due to competition for resources, at rate $1/N$. The foxes die or leave with a constant per-capita rate, and the units of time are chosen such that this rate is equal to $1$. The parameters $s$ and $\tau$ are related to the predation rate as follows: For a small rabbit population $R$, the predation rate, $sR$, is proportional to $R$. For a very large rabbit population, the predation rate saturates at $1/\tau$ so as to describe satiation of the predators.

\begin{table}[ht]
\centering % used for centering table
\begin{tabular}{|c c c|} % centered columns (3 columns)
\hline\hline %inserts double horizontal lines
Process & Type of Transition & Rate \\ [0.5ex] % inserts table
%heading
\hline % inserts single horizontal line
Birth of rabbits & $R \rightarrow 2R$ 				& $aR$ \\ % inserting body of the table
Predation and birth of foxes & $F+R \rightarrow 2F$	& $\frac{sRF}{1+s\tau R}$ \\ % inserting body of the table
Death of foxes & $F \rightarrow 0$ 					& $F$ \\ % inserting body of the table
Competition among rabbits & $2R \rightarrow R$ 			& $\frac{R\left(R+1\right)}{2N}$  \\ [1ex] % [1ex] adds vertical space
\hline %inserts single line
\end{tabular}
\caption{Stochastic  Rosenzweig-MacArthur  model.}
\label{table:reactions} % is used to refer this table in the text
\end{table}

The deterministic equations for the RMA model are:
 \begin{eqnarray}
\label{eq_mean_field_orig}
\dot{R}&=&aR-\frac{1}{2N}R^{2}-\frac{sRF}{1+s\tau R} \\
\label{eq_mean_field_orig2}
\dot{F}&=&-F+\frac{sRF}{1+s\tau R} .
 \end{eqnarray}
We assume that $s$ scales with the system size as $s\propto 1/N$, where $N\gg1$ is the scale of the population sizes.
We introduce $x= R/N$, $y= F/N$, and  $\sigma= sN ={\mathcal O} (1)$ and arrive at the rescaled equations
\begin{equation}
\label{eq_xydot_mean_field}
\dot{x}=ax-\frac{1}{2}x^{2}-\frac{\sigma xy}{1+\sigma\tau x}, \quad \dot{y}=-y+\frac{\sigma xy}{1+\sigma\tau x} ,
\end{equation}
where all the quantities are assumed to be of order $1$.

\begin{figure}
\includegraphics[width=2.5 in,clip=]{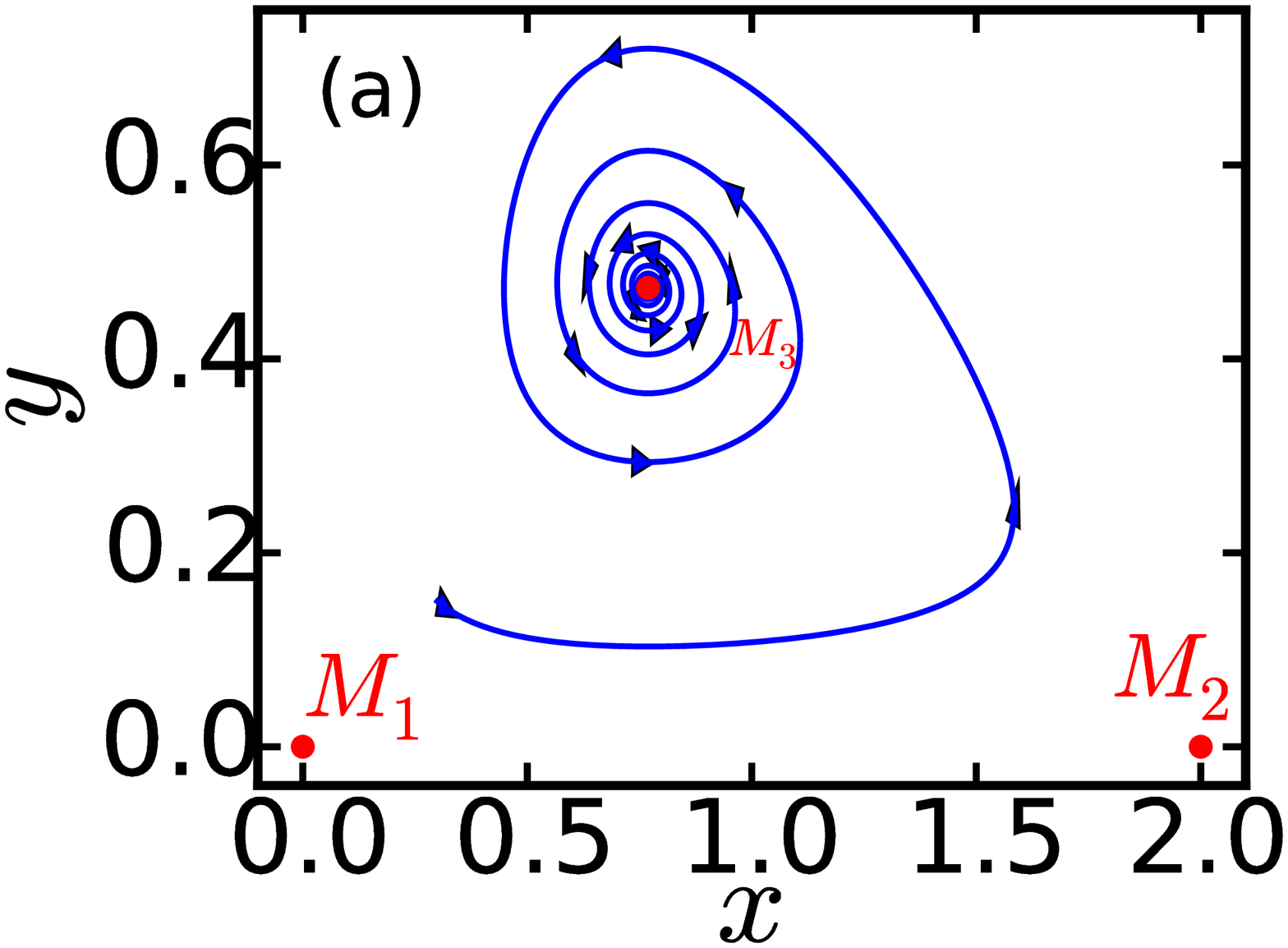}
\includegraphics[width=2.5 in,clip=]{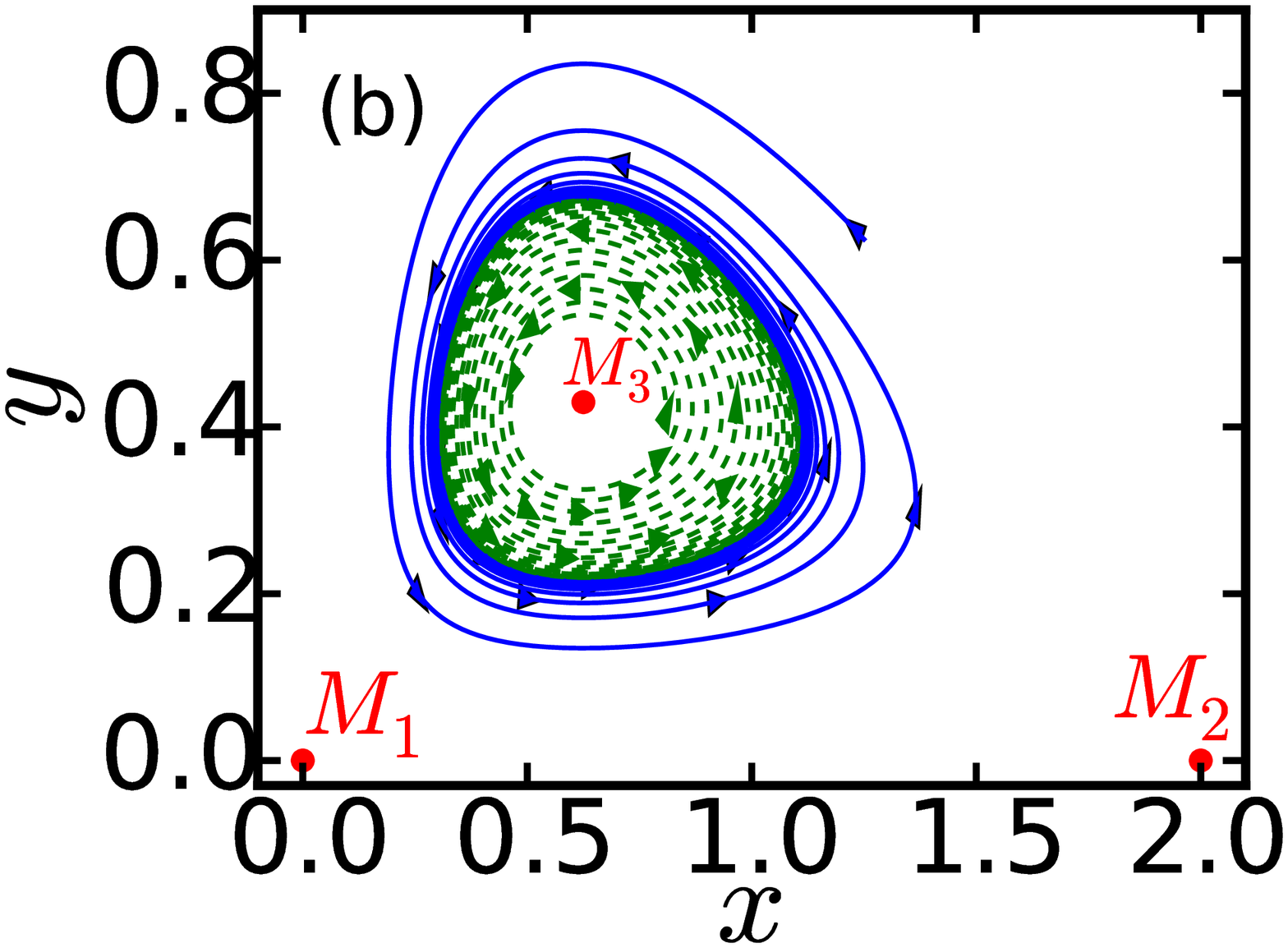}
\caption{(Color online) Phase portrait of the deterministic RMA model (\ref{eq_xydot_mean_field}) for $a=1$, $\tau=0.5$ and
two different values of $\sigma$: $\sigma=2.6$, where $M_3$ is a stable focus (a), and $\sigma=3.2$, where $M_3$ is an unstable focus (b).}
\label{fig_deterministic}
\end{figure}

The deterministic RMA model has been extensively studied \citep{RM1963,Cheng1981,HalSmith2008}. Here we recap the main results of these works that we need for our purposes.
We are interested only in the regime of parameters
\begin{equation}
\label{eq_tau_and_sigma_condition}
0<\tau< 1,\quad  \sigma>\sigma_0=\frac{1}{2a\left(1-\tau\right)} ,
\end{equation}
when Eqs.~(\ref{eq_xydot_mean_field}) have three fixed points describing nonnegative population sizes.
The fixed point $M_1$ $\left(\bar{x}_{1}=0,\bar{y}_{1}=0\right)$ corresponds to an empty system. It is a saddle point: attracting in the $y$-direction (when there are no rabbits in the system), and repelling in the $x$-direction. The fixed point $M_2$ $\left(\bar{x}_{2}=2a,\bar{y}_{1}=0\right)$ describes a steady-state population of rabbits in the absence of foxes. It is also a saddle: attracting in the $x$-direction (when there are no foxes), and repelling in a direction corresponding to the introduction of a few foxes into the system. The third fixed point $M_3$ $\left(\bar{x}_{3},\bar{y}_{3}\right)$, where
\begin{equation}
\label{eq_M3_coordinates}
\bar{x}_3=\frac{1}{\sigma\left(1-\tau\right)}, \quad \bar{y}_3=\frac{2a\sigma\left(1-\tau\right)-1}{2\sigma^{2}\left(1-\tau\right)^{2}},
 \end{equation}
describes the coexistence state of the rabbits and foxes. Its stability properties depend on the parameters: For
\begin{equation}
\label{eq_sigma_node2focus}
\sigma_0 < \sigma < \bar{\sigma}=\frac{\frac{a\tau\left(1+\tau\right)}{2\left(1-\tau\right)}-1-\sqrt{1+a\frac{1+\tau}{1-\tau}}}{a^{2}\tau^{2}-4a\left(1-\tau\right)} ,
 \end{equation}
$M_3$ is a stable node. For
\begin{equation}
\label{eq_sigma_star}
\bar{\sigma} < \sigma < \sigma^{*}=\frac{1+\tau}{2a\tau\left(1-\tau\right)}
\end{equation}
it is a stable focus. Finally, for $\sigma > \sigma^{*}$, $M_3$ is unstable, and a stable limit cycle appears around it. Noise-driven population extinction from a limit cycle has not been studied before. Therefore,  in most of the paper we assume that  $\sigma > \sigma^{*}$. A Hopf bifurcation occurs at
$\sigma=\sigma^{*}$. Figure \ref{fig_deterministic} shows the the behaviors of the deterministic model for $\bar{\sigma} < \sigma < \sigma^{*}$ (a) and for  $\sigma > \sigma^{*}$ (b). The characteristic time scale $t_r$ of the deterministic dynamics is determined by the real part of the eigenvalues of the linear stability matrix at the fixed point $M_3$ when the latter is stable, and by the period of the limit cycle and the relaxation time toward it when $M_3$ is unstable.

\section{Stochastic Rosenzweig-MacArthur Model: Extinction from a Limit Cycle}
\label{sec:stochastic}

\subsection{Two routes to extinction}
\label{sec:stochastic:two}

\begin{figure}
\includegraphics[width=3.4 in,clip=]{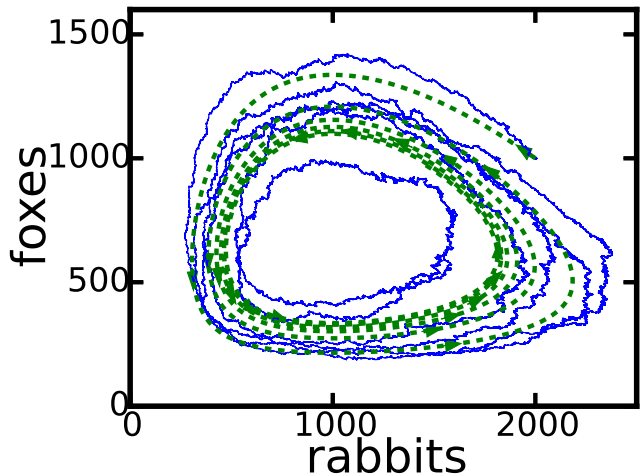}
\caption{(Color online) Phase space trajectories of the deterministic RMA model (dashed) vs. a Monte-Carlo simulation of the stochastic model (solid). The parameters are $a=1, \sigma=3.2, \tau=0.5$, and $N=1600$. For intermediate times, the stochastic dynamics closely follow the deterministic dynamics. For much longer times (not shown), at least one of the populations goes extinct, see Fig. \ref{fig_monte_carlo_oscillations}.
}
\label{fig_montecarlo1}
\end{figure}

\begin{figure}
\includegraphics[width=3.5 in,clip=]{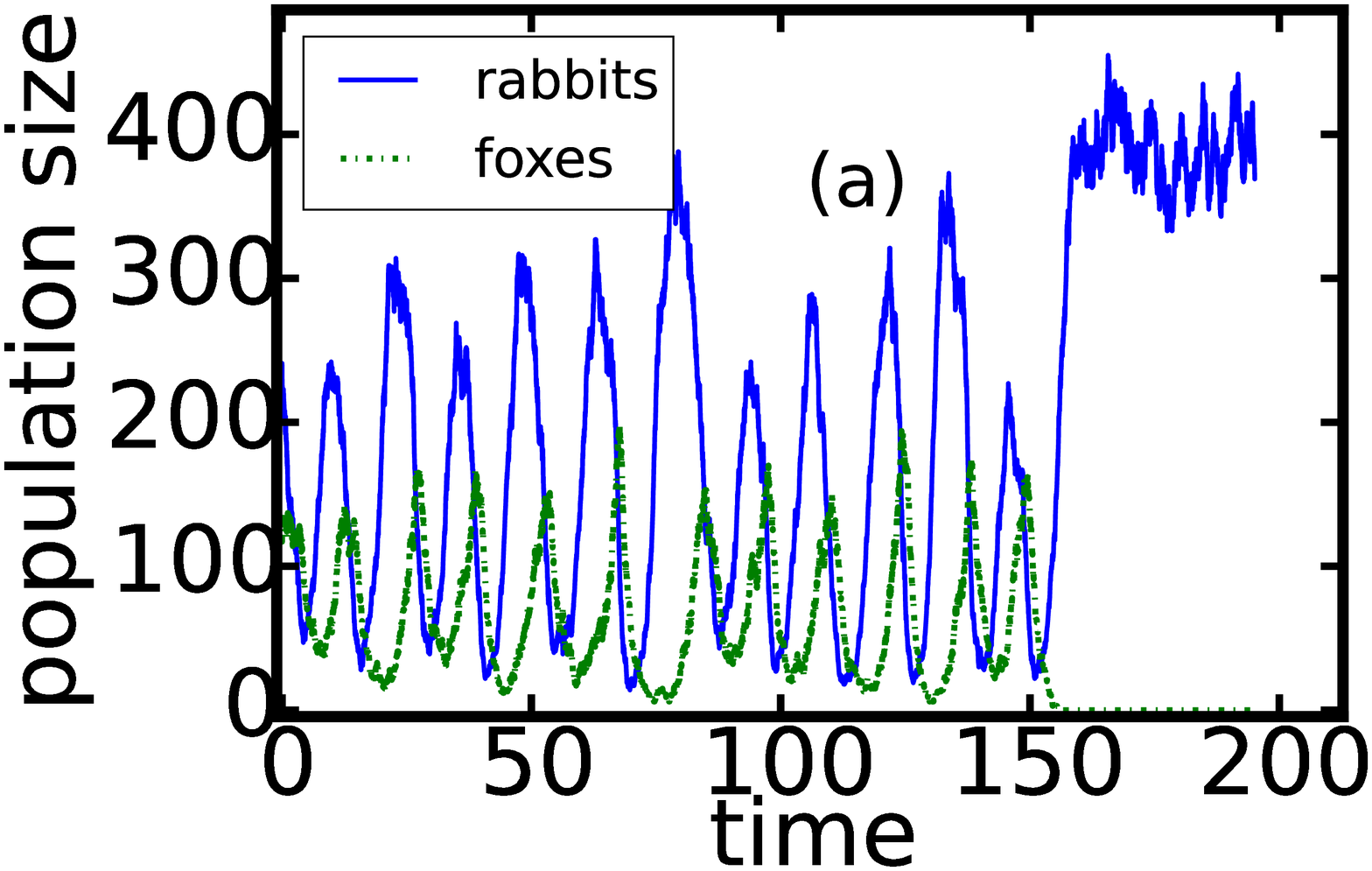}
\includegraphics[width=3.5 in,clip=]{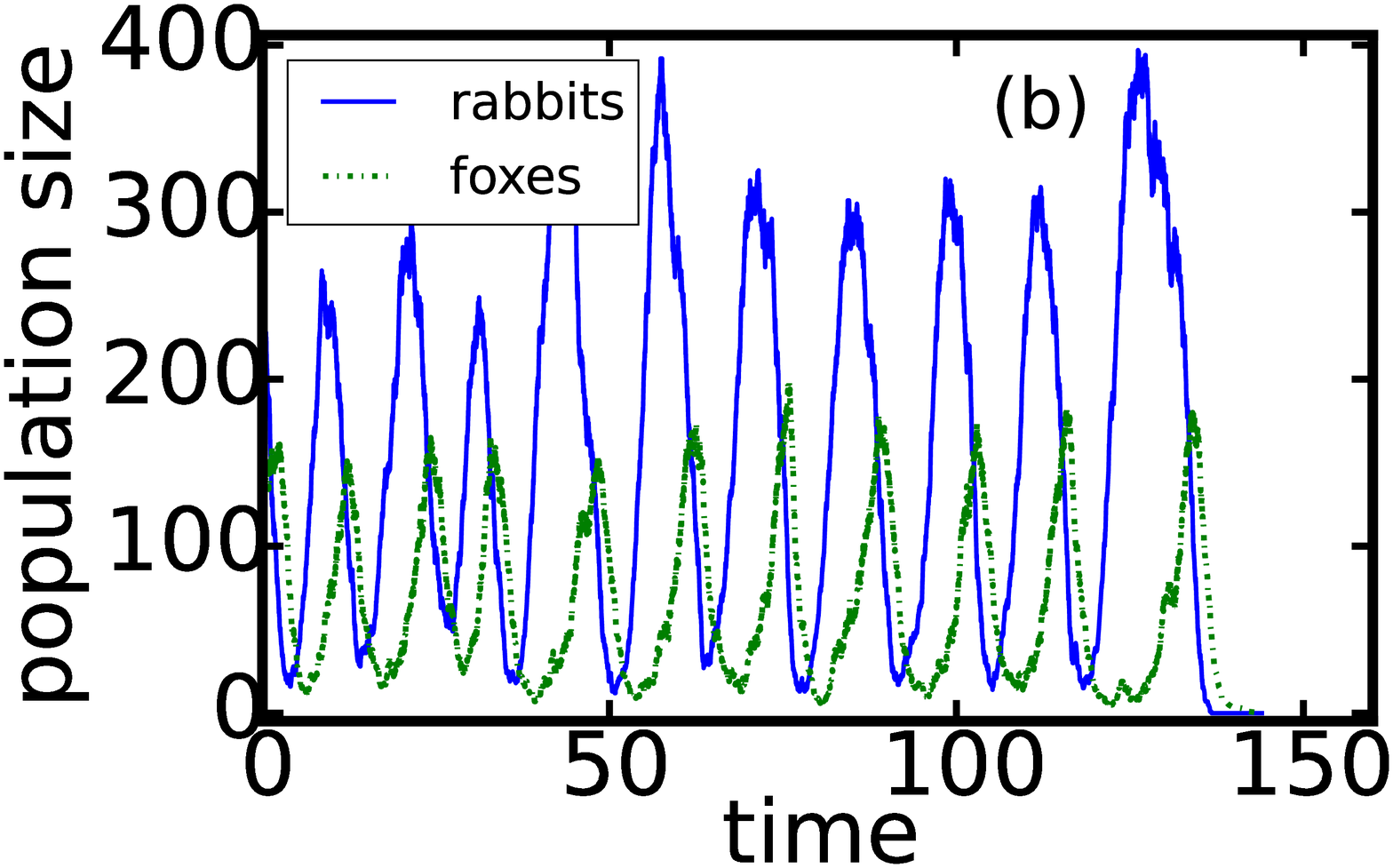}
\caption{(Color online) Two stochastic realizations of the RMA model for $a=1$, $\tau=0.5$, $\sigma=3.2$, and $N=192$. Shown are the population sizes of rabbits (solid lines) and foxes (dashed lines) versus time.  (a) The foxes go extinct first, whereas the rabbits approach a steady state around the fixed point $M_2$, and live on forever. (b) The rabbits go extinct first, at $t\simeq137$, and the foxes go extinct very soon afterwards, at $t\simeq142$.}
\label{fig_monte_carlo_oscillations}
\end{figure}

How does the deterministic picture change when one accounts for the stochasticity of the elemental processes of the RMA model and the discrete character of the population sizes?
Figure \ref{fig_montecarlo1} shows a stochastic realization of the RMA model in the $R,F$ plane. As one can see, at sufficiently large $N$, and at intermediate times, the stochastic trajectory closely follows the deterministic one.  The long-time behavior, however, is dramatically different: At least one of the populations here goes extinct, and this happens in one of two possible ways. Figure \ref{fig_monte_carlo_oscillations} (a) and (b) shows  two different stochastic realizations for the same values of parameters (which coincide with those in Fig. \ref{fig_deterministic}b), and for the same initial conditions. In figure a the foxes go extinct, while the rabbit population approaches a nonzero steady state. In figure b the rabbits go extinct first, followed by a quick extinction of the foxes. The population extinction results from the presence of two absorbing states in this system: the empty state, and the state without foxes but with a nonzero rabbit population (Note that, in our model, the rabbits are immortal in the absence of foxes.).  One of these two absorbing states is always ultimately reached. At $N\gg 1$ this usually happens due to a rare large fluctuation when starting from the long-lived population state in the vicinity of the deterministic limit cycle.

\subsection{Master equation and long-time dynamics}
\label{sec:stochastic:master}

Let  $P_{m,n}\left(t\right)$ be the probability to find $m$ rabbits and  $n$ foxes in the system at time $t$. The dynamics of $P_{m,n}\left(t\right)$ is governed by the master equation:
\begin{eqnarray}
\label{MasterEquation}
    \dot{P}_{m,n}&\!=\!&\hat{H} P_{m,n}=a\left[\left(m-1\right)P_{m-1,n}-mP_{m,n}\right]+\nonumber \\
    &\!+\!&\frac{\sigma\left(m+1\right)\left(n-1\right)}{N+\sigma\tau\left(m+1\right)}P_{m+1,n-1}-\frac{\sigma mn}{N+\sigma\tau m}P_{m,n}+\nonumber \\
    &\!+\!&\left(n+1\right)P_{m,n+1}-nP_{m,n}+\nonumber \\
    &\!+\!&(1/2N)[(m+1)m P_{m+1,n}-m(m-1)P_{m,n}],
\end{eqnarray}
where $P_{m,n}=0$ when any of the indices is negative. At times much longer than $t_r$ but shorter than the MTE (see below), a quasistationary distribution appears around the deterministic limit cycle. %\citep{Movie}. 
Following Ref.~\citep{GM2012}, we can define the ``effective probability
contents" of the vicinities of the fixed points $M_1$, $M_2$ and the limit cycle:
\begin{eqnarray}
  \mathcal{P}_1 (t) &=& P_{0,0} (t)\,, \label{calP1} \\
  \mathcal{P}_2 (t) &=& \sum_{m=1}^{\infty}P_{m,0} (t)\,, \label{calP2} \\
  \mathcal{P}_3 (t) &=& \sum_{m=1}^{\infty} \sum_{n=1}^{\infty} P_{m,n} (t)\,. \label{calP3}
\end{eqnarray}
Assuming $N\gg 1$ and $t\gg t_r$, the main contributions to the sums in Eqs.~(\ref{calP2}) and (\ref{calP3}) come from the close vicinities of the fixed point $M_2$ and the limit cycle, respectively. The effective long-times dynamics of $\mathcal{P}_1$,  $\mathcal{P}_2$ and  $\mathcal{P}_3$  are given by a three-state master equation \citep{GM2012}:
\begin{eqnarray}
\label{efrateeqn}
  \dot{\mathcal{P}}_1 (t) &=& \mathcal{R}_1 \mathcal{P}_3 (t)\,,\nonumber \\
  \dot{\mathcal{P}}_2 (t)&=&  \mathcal{R}_2 \mathcal{P}_3 (t)\,,\nonumber \\
  \dot{\mathcal{P}}_3 (t)&=& -(\mathcal{R}_1 +\mathcal{R}_2) \mathcal{P}_3 (t) \,,
\end{eqnarray}
where $\mathcal{R}_{1}$ and $\mathcal{R}_{2}$ are the extinction rates along the first and second extinction route, respectively. At $t\gg t_r$, We assume that initially the populations
occupy the coexistence state in the vicinity of the limit cycle:
\begin{equation}\label{incond}
   \left[\mathcal{P}_{1}\left(0\right),\,\mathcal{P}_{2}\left(0\right),\,\mathcal{P}_{3}\left(0\right)\right]=\left(0,0,1\right)\,.
\end{equation}
Then the solution of Eqs. ~(\ref{efrateeqn}) is \citep{GM2012}
\begin{eqnarray}
\mathcal{P}_1 (t) &=& \frac{\mathcal{R}_{1}\,\left[1-e^{-\left(\mathcal{R}_{1}+\mathcal{R}_{2}\right)t}\right]}{\mathcal{R}_{1}+\mathcal{R}_2},\label{P1}\\
\mathcal{P}_2 (t) &=&\frac{\mathcal{R}_{2}\,\left[1-e^{-\left(\mathcal{R}_{1}+\mathcal{R}_{2}\right)t}\right]}{\mathcal{R}_{1}+\mathcal{R}_2},  \label{P2} \\
\mathcal{P}_3 (t)  &=& e^{-(\mathcal{R}_1 +\mathcal{R}_2)t}.\label{P3}
\end{eqnarray}
As one can see from Eqs.~(\ref{P1})-(\ref{P3}), the long-term behavior of the system is determined by the extinction rates $\mathcal{R}_{1}$ and $\mathcal{R}_{2}$.
Their evaluation, therefore, will be our main objective.

From Eqs.~(\ref{P1})-(\ref{P3}), the mean time to extinction of foxes is \citep{GM2012}
\begin{eqnarray}
\label{MTEfox}
  \text{MTE}_F&=& \int_{0}^{\infty}\!\!dt\, t\!\left[\dot{P}_{1}\left(t\right)+\dot{P}_{2}\left(t\right)\right]=-\int_{0}^{\infty}\!\!dt\, t\dot{P}_{3}\left(t\right)=\nonumber \\
  &=&1/\left(\mathcal{R}_{1}+\mathcal{R}_{2}\right),
\end{eqnarray}
whereas the mean time to extinction of the rabbits is formally infinite. As we shall see below, the first extinction scenario is much less likely than the second, i.e. $\mathcal{R}_1 \ll \mathcal{R}_2$. As a result, the MTE of the foxes can be approximated as
\begin{equation}
\label{MTEfox_approx}
  \text{MTE}_F\simeq 1/\mathcal{R}_2 .
\end{equation}

We now briefly discuss, for completeness, how the extinction rates are encoded in the spectrum of the linear operator $\hat{H}$, introduced in Eq.~(\ref{MasterEquation}). Let us denote the eigenvalues and eigenstates of $\hat{H}$ by $\lambda_i$ and $\pi_{m,n}^{\left(i\right)}$, respectively, i.e.
\begin{equation}
\label{full}
\hat{H}\pi_{m,n}^{\left(i\right)}=-\lambda_{i}\pi_{m,n}^{\left(i\right)}.
\end{equation}
We can write the solution of the time-dependent master equation (\ref{MasterEquation}) in terms of the eigenstates as
\begin{equation}
P_{m,n}\left(t\right)=\sum_{i=1}^{\infty}C_{i}\pi_{m,n}^{\left(i\right)}e^{-\lambda_{i}t} ,
\end{equation}
where the constants $C_{i}$ are determined by the initial condition $P_{m,n}\left(0\right)$ \citep{GM2012}.
Two of the eigenstates describe the (truly) steady-state solutions, corresponding to the empty state, and the fox-free state with a steady-state population of rabbits. Their corresponding eigenvalues are both equal to zero. The smallest nonzero eigenvalue is $\lambda_3 = \mathcal{R}_1 + \mathcal{R}_2$ \citep{GM2012}.

Our principal task in the remainder of this paper is to evaluate this eigenvalue and the extinction rates $\mathcal{R}_1$ and $\mathcal{R}_2$, by examining the eigenvalue problem
\begin{equation}\label{QSD_eq0}
    \hat{H}\pi_{m,n}=-\left(\mathcal{R}_{1}+\mathcal{R}_{2}\right) \pi_{m,n}\quad n>0 .
\end{equation}
Since the extinction rates are exponentially small in $N\gg1$, one can approximate the full Eq.~(\ref{full}) for $\pi_{m,n}$ by the quasi-stationary equation
\begin{equation}\label{QSD_eq}
    \hat{H}\pi_{m,n}\simeq 0, \quad m>0,\;n>0\,.
\end{equation}

\subsection{WKB approximation}
\label{sec:stochastic:WKB}

\subsubsection{General}

For $N\gg 1$, Eq.~(\ref{QSD_eq}) can be approximately solved via the WKB ansatz \citep{Dykman1994}
\begin{equation}\label{WKB_ansatz}
    \pi_{m,n}=\exp[-N S(x,y)]\,,
\end{equation}
where $x=m/N$ and $y=n/N$. Assuming that $S\left(x,y\right)$ is a smooth function of $x$ and $y$, we plug this ansatz  into Eq.~(\ref{QSD_eq}) and Taylor-expand $S$ around $\left(x,y\right)$.   In the leading order in $1/N$, this procedure yields a zero-energy Hamilton-Jacobi equation $H(x,y,\partial_xS,\partial_yS)=0$, with the effective Hamiltonian
\begin{eqnarray}\label{Hamiltonian}
   H\left(x,y,p_{x},p_{y}\right)\!\!&=&\!\!ax\left(e^{p_{x}}-1\right)+\frac{\sigma xy}{1+\sigma\tau x}\left(e^{p_{y}-p_{x}}-1\right)+\nonumber\\
   &+&\!\!y\left(e^{-p_{y}}-1\right)+\frac{x^{2}}{2}\left(e^{-p_{x}}-1\right).
\end{eqnarray}
The Hamilton equations are
\begin{eqnarray}
\label{Motion}
\dot{x}\!&=&\!axe^{p_{x}}-\frac{x^{2}}{2}e^{-p_{x}}-\frac{\sigma xy}{1+\sigma\tau x}e^{p_{y}-p_{x}},\nonumber \\
\dot{y}\!&=&\!\frac{\sigma xy}{1+\sigma\tau x}e^{p_{y}-p_{x}}-ye^{-p_{y}} ,\nonumber \\
\dot{p}_{x}\!&=&\!a\left(1\!-\!e^{p_{x}}\right)\!+\!x\left(1\!-\!e^{-p_{x}}\right)\!+\!\frac{\sigma y}{\left(1\!+\!\sigma\tau x\right)^{2}}\left(1\!-\!e^{p_{y}-p_{x}}\right) ,\nonumber \\
\dot{p}_{y}\!&=&\!\left(1-e^{-p_{y}}\right)+\left(1-e^{p_{y}-p_{x}}\right)\frac{\sigma x}{1+\sigma\tau x} .
\end{eqnarray}
We are only interested in the zero-energy manifold $H=0$. Note that in the zero-energy invariant hyperplane $p_x=p_y=0$, the dynamics of $x$ and $y$ reduce to the deterministic dynamics (\ref{eq_xydot_mean_field}).
Therefore, the three deterministic fixed points $M_{1}=\left(0,0,0,0\right)\; M_{2}=\left(2a,0,0,0\right)\; M_{3}=\left(x^{*},y^{*},0,0\right)$ are also fixed points of the Hamiltonian dynamics (\ref{Motion}). In its turn, the deterministic limit cycle is an exact time-periodic solution of the Hamilton equations (\ref{Motion}) with $p_x=p_y=0$.
In addition, there are two ``fluctuational" fixed points:
\begin{eqnarray}
  F_{1}&=& \left(0,0,0,-\infty\right),  \nonumber\\
  F_{2}&=&\left[2a,0,0,\ln\left(\frac{1+2a\tau\sigma}{2a\sigma}\right)\right] .
  \label{FFpoints}
\end{eqnarray}
Fluctuational fixed points have a non-zero $p_x$ or $p_y$ component and appear in a whole class of stochastic population models that exhibit extinction in the absence of an Allee effect \cite{EK2,Dykman2008,Kamenev2008,OvMe2010,Assaf2010metastable}.

We can evaluate the extinction rates $\mathcal{R}_{1}$ and $\mathcal{R}_{2}$ by evaluating the QSD $\pi_{m,n}$ at $\left(x=0,y=0\right)$ and $\left(x=2a,y=0\right)$, respectively. This is done by calculating the actions $S_1$ and $S_2$ along the corresponding instantons: phase space trajectories which begin, at $t=-\infty$, on the deterministic limit cycle, and end, at $t=\infty$, at the fluctuational fixed point $F_1$ or $F_2$, respectively.
These actions
$$
S_{1}\!=\!\int_{\left(x_{lc},y_{lc}\right)}^{F_{1}}\!\!\!\!\!\!\!\!p_{x}dx+p_{y}dy\quad\mbox{and}\quad S_{2}\!=\!\int_{\left(x_{lc},y_{lc}\right)}^{F_{2}}\!\!\!\!\!\!\!\!p_{x}dx+p_{y}dy,
$$
With exponential accuracy, the extinction rates $\mathcal{R}_{1}$ and $\mathcal{R}_{2}$ are \citep{GM2012}
\begin{equation}\label{rates}
  \mathcal{R}_{1} \sim  \exp(-NS_{1}), \quad  \mathcal{R}_{2} \sim  \exp(-NS_{2})\,.
\end{equation}

\subsubsection{Some properties of the instantons. Shooting method}
\label{section_floquet_theory}

For the case of extinction from a fixed point, the instantons can be found numerically: either by shooting \cite{Kamenev2008,Dykman2008,GM2012}, or by iterating the equations for $\dot{x}$ and $\dot{y}$ forward in time, and equations for $\dot{p}_x$ and $\dot{p}_y$ backward in time \cite{EK2,Stepanov,Lohmar2011}.

Here we modify the shooting method \cite{Kamenev2008,Dykman2008,GM2012} to make it suitable for an instanton which exits, at $t=-\infty$, a deterministic limit cycle and enters, at $t=\infty$, one of the fluctuational fixed points. First we need to discuss some important properties of such instantons. In particular, how the instanton exits the limit cycle is crucial for the shooting method we are about to present.

In the case of extinction from an attracting fixed point, one proceeds by linearizing the Hamilton equations near the fixed point, and finding the unstable eigenvectors of the linearizing matrix. The matrix will have two ``stable" (deterministic) eigenvectors, and two ``unstable" eigenvectors, whose eigenvalues have a positive real part, see e.g. Eq. (7.25) in Ref. \citep{ChaosBook}. The instanton (which, in this case, is a heteroclinic trajectory going to a fluctuational fixed point) is then found by looking for a linear combination of the two unstable eigenvectors of a fixed (and very small) norm, using the shooting method \citep{Dykman2008,Kamenev2008}.

When the extinction is from a stable limit cycle, the leading-order dynamics in the vicinity of the limit cycle is best described by Floquet theory, see e.g. \citep{Ward2010}. For each point $\vect{v_{lc}}=\left(x_{lc},y_{lc}, 0, 0\right)$ on the limit cycle, we define its Floquet matrix $\mat{B}$ as follows: Given a starting point which is near the limit cycle, $\vect{v_{lc}}+ \vect{\delta v}$, and advancing in time according to the Hamilton equations, we will arrive, after one period of the limit cycle, at the point $\vect{v_{lc}}+ \mat{B} \vect{\delta v} +O\left(\left\Vert \vect{\delta v}\right\Vert ^{2}\right)$. The eigenvectors and eigenvalues of the Floquet matrix will determine the stable and unstable directions in phase space.  (We assume everywhere in the following that the eigenvectors are normalized to unity.) Although the Floquet matrix $\mat{B}$ itself and its eigenvectors are local quantities which vary along the limit cycle, its eigenvalues are independent of the choice of the point $\left(x_{lc},y_{lc}, 0, 0\right)$ on the limit cycle \citep{Ward2010,ChaosBook}.

As in the fixed point case, $\mat{B}$ will have two ``deterministic" (zero-momentum) eigenvectors. The corresponding eigenvalues will be $\lambda$ and $1$. $\lambda<1$ corresponds to a stable eigenvector, and $1$ corresponds to the neutral eigenvector, which is tangent to the limit cycle at its every point. Note that $\lambda$ must be positive, otherwise phase-space trajectories would cross each other. The two additional eigenvalues of $\mat{B}$ must be equal to $1$ (another neutral eigenvector) and $1/\lambda$ (an unstable eigenvector), see Eq.~(7.27) of Ref. \citep{ChaosBook} or Ref. \cite{masterthesis} for the proof. Therefore, there is only \emph{one} unstable direction through which a trajectory can exit the limit cycle. We now understand how the instanton exits the limit cycle. It starts on the limit cycle at $t= -\infty$, and then performs an infinite number of rotations around the limit cycle, each one further from the limit cycle by a factor of $1/\lambda$, before departing.

In view of this basic property of the instanton, our numerical method consists of several steps. The first step is to choose an arbitrary point $\vect{v_{lc}}=\left(x_{lc},y_{lc}, 0, 0\right)$ on the limit cycle. This is done by numerically integrating the deterministic equations~(\ref{eq_xydot_mean_field}). The period $T$ of the limit cycle is computed numerically as well.

We then numerically compute the Floquet matrix $\mat{B}$ at the point $\vect{v_{lc}}$ we chose. This is done by adding small perturbations to each of the 4 coordinates of $\vect{v_{lc}}$ in turn, and then advancing the Hamiltonian equations~(\ref{Motion}) from time $t=0$ to $t=T$. The result is a point which is near $\vect{v_{lc}}$, but the small distance from it gives us a column of the Floquet matrix. For example: We set $\vect{u}\left(t=0\right)=\left(x_{lc},y_{lc},\delta p_{x},0\right)$, and then advance $\vect{u}\left(t\right)$ according to the Hamilton equations, until time $t=T$. The third column of the Floquet matrix is then given by $\left[\vect{u}\left(t=T\right)-\vect{v_{lc}}\right]/\delta p_{x} + O\left(\delta p_{x}\right)$.

Next we diagonalize the Floquet matrix, and find its only unstable vector $\vect{v}$, whose eigenvalue $1/\lambda$ is larger than $1$. Since the instanton must exit the limit cycle through this unstable direction, we expect there to be a discrete set of $\epsilon$'s for which $\vect{v_{lc}}+\epsilon \vect{v}$ is on the instanton (to leading order in $\epsilon$). What is left is to find one such $\epsilon$ by the shooting method. We emphasize that $\epsilon$ can be taken to be as small as we like, because if $\vect{v_{lc}}+\epsilon \vect{v}$ is on the instanton, then so is $\vect{v_{lc}}+\lambda \epsilon \vect{v}$ to leading order in $\epsilon$.

It is easy to show that $\partial H/\partial v \equiv \nabla H \cdot \vect{v}$,  the derivative of the Hamiltonian $H$ in the direction of the unstable eigenvector $\vect{v}$ vanishes.
Let us start from some initial condition:
\begin{equation*}
 \vect{u}\left(t=0\right) = \vect{u_0} = \left(x_{lc},y_{lc}, 0, 0\right) + \epsilon \vect{v}.
\end{equation*}
Assuming that $\epsilon$ is small enough, $\epsilon \ll 1$, the energy at this point is $H\left(u_{0}\right)\simeq\epsilon\,\partial H/\partial v$. After advancing the Hamiltonian dynamics by the period $T$ of the limit cycle, the system will be at the point
\begin{equation*}
\vect{u}\left(t=T\right)\simeq\left(x_{lc},y_{lc},0,0\right)+\lambda^{-1} \epsilon \vect{v},
\end{equation*}
and the energy will be
\begin{equation*}
H\left[\vect{u}\left(t=T\right)\right]\simeq\lambda^{-1} \epsilon\frac{\partial H}{\partial \vect{v}}.
\end{equation*}
By virtue of energy conservation,
\begin{equation*}
H\left[\vect{u}\left(t=T\right)\right]=H\left(\vect{u_{0}}\right).
\end{equation*}
Since $\lambda$ is different from $1$ ($\lambda<1$), this implies that $\partial H/\partial \vect{v} = 0$.

As we see now, there are three directions which are tangent to the zero-energy manifold -- the two deterministic directions and the unstable direction $\vect{v}$. Is the fourth direction -- the neutral ``quantum" eigenvector -- also tangent to the zero-energy manifold? The answer is no. The reason is that, on the limit cycle, $\nabla H \ne 0$ (the gradient of $H$ vanishes only at fixed points of the Hamiltonian dynamics), so there cannot be four independent directions for which the directional derivative is zero.

\subsubsection{Finding the instantons and evaluating $S_{1}$ and $S_{2}$}
\label{sec:stochastic:instantons:numerical}

\begin{figure}
\includegraphics[width=3.3 in,clip=]{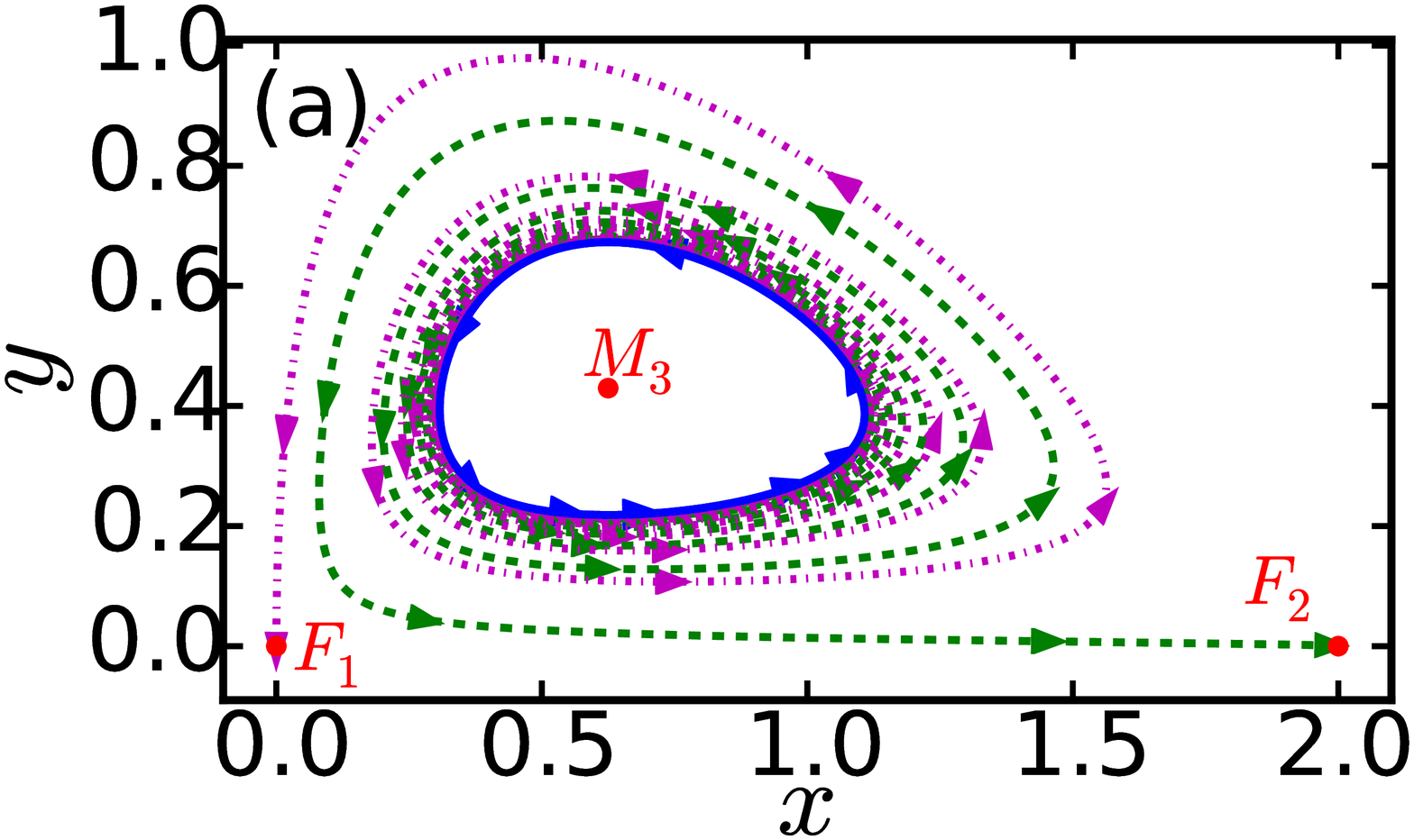}
\includegraphics[width=2.8 in,clip=]{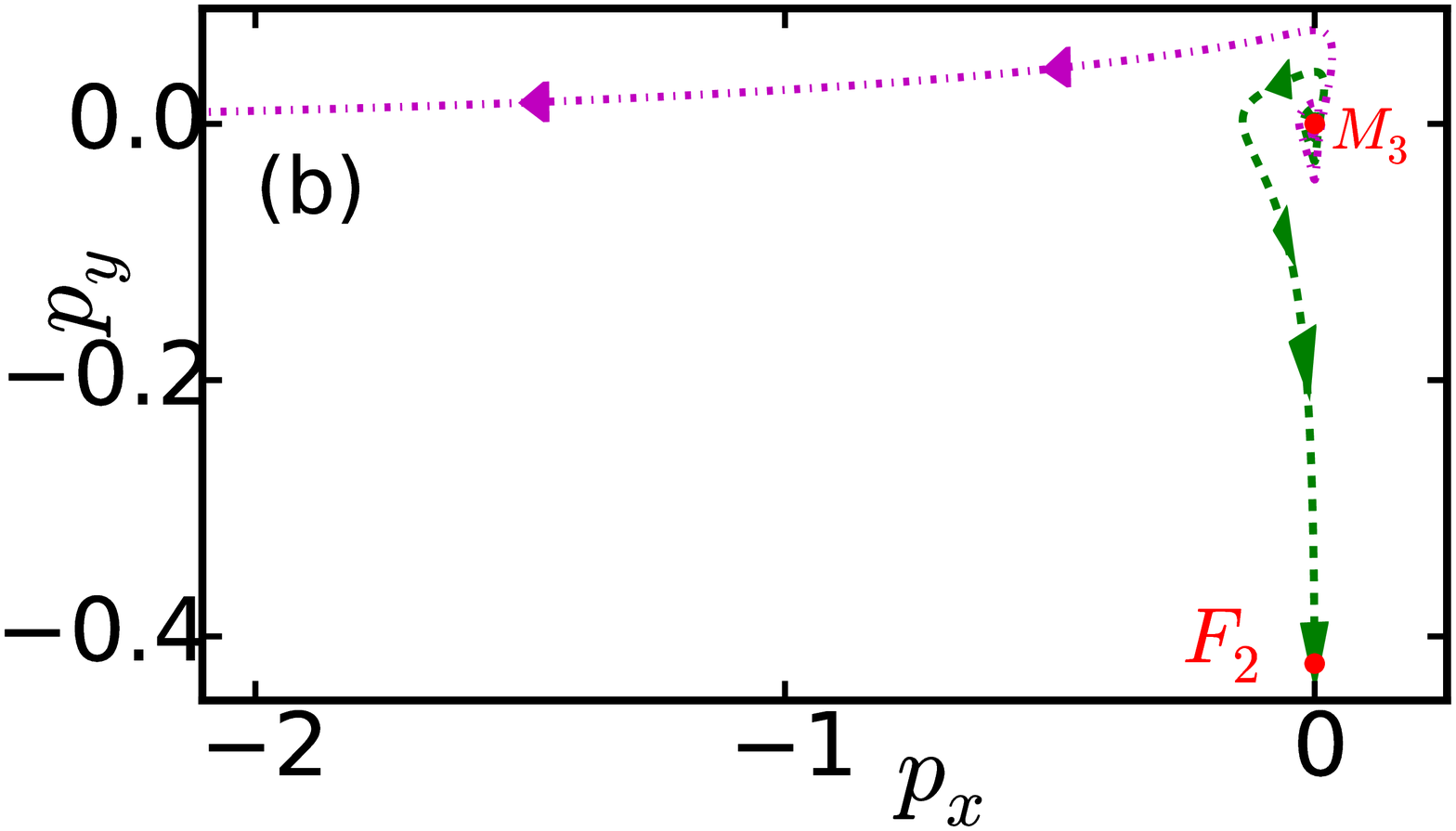}
\caption{(Color online) Numerically found instantons from the stable limit cycle (solid line) to $F_1$ (dot-dashed) and to $F_2$ (dashed). The parameters are $a=1,\;\sigma=3.2$, and $\tau=0.5$. Shown are
the $xy$-projections (a) and the $p_xp_y$ projections (b).}
\label{nodeinst}
\end{figure}

\begin{figure}
\includegraphics[width=3.0 in,clip=]{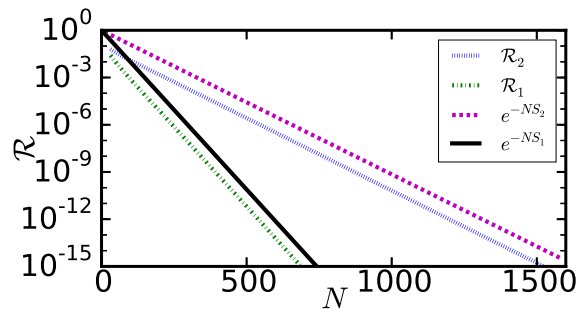}
\caption{(Color online) Extinction rates  $\mathcal{R}_{i}$, determined by solving the master equation (\ref{MasterEquation})
numerically for different $N$, are compared with the extinction rates $e^{-NS_{i}}$ calculated in the leading-order WKB approximation. The parameters are $a=1$, $\tau=0.5$, and $\sigma=3.1$. The actions along the instantons are $S_1 \simeq 0.0466$ and $S_2 \simeq 0.0211$. The vertical shifts are due to the undetermined WKB prefactors $\mathcal{F}_1$ and $\mathcal{F}_2$, see Sec. \ref{sec:prefactor}.}

\label{fig_extinction_rates}
\end{figure}

\begin{figure}
\includegraphics[width=3.5 in,clip=]{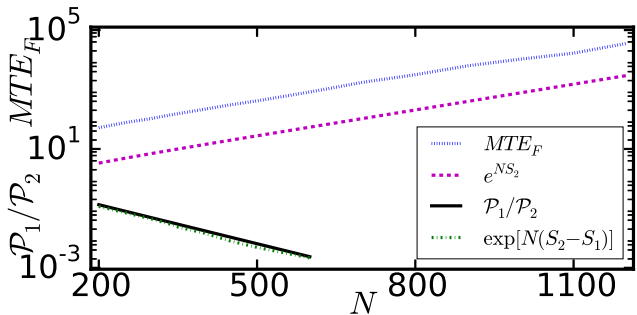}
\caption{(Color online) $N$-dependence of the mean time to extinction of the foxes $\text{MTE}_{F}$ and of the ratio of probabilities of the two extinction scenarios $\mathcal{P}_{1}/\mathcal{P}_{2}$ for fixed $a=1$, $\tau=0.5$, and $\sigma=4$.
The results obtained by averaging over many Monte-Carlo simulations are compared with the leading-order WKB predictions. The values are plotted on a semi-log scale. The actions are $S_1 \simeq 0.0167$ and $S_2 \simeq 0.00665$.  Since $S_2 < S_1$, the MTE of the foxes is approximated in the WKB theory as $\text{MTE}_{F}\simeq e^{NS_{2}}$, see Eq.~(\ref{MTEfox_approx}).
The vertical shifts are due to the undetermined WKB prefactors $\mathcal{F}_{i}$, see Sec. \ref{sec:prefactor}.}
\label{fig_MTE_and_P1_div_P2}
\end{figure}

Examples of numerically found instantons which start at the limit cycle and end at the fluctuational fixed points  $F_1$ and $F_2$ are shown in Fig. \ref{nodeinst}.
Figure \ref{fig_extinction_rates} compares the extinction rates $\mathcal{R}_{i}$ obtained by solving numerically the (truncated) master equation (\ref{MasterEquation}), with the result of the leading order WKB approximation, $e^{-NS_i}$. Figure \ref{fig_MTE_and_P1_div_P2}  shows our results for the the mean time to extinction of foxes $\text{MTE}_{F}$ and for the ratio of the probabilities of the two extinction routes, $\mathcal{P}_{1}/\mathcal{P}_{2}$, obtained by averaging over many Monte-Carlo simulations. The same figure shows the corresponding leading-order WKB predictions  $1/\left(e^{-NS_{1}}+e^{-NS_{2}}\right)\simeq e^{NS_{2}}$ and $e^{N\left(S_{2}-S_{1}\right)}$. In both cases a good agreement, up to undetermined WKB prefactors, is observed between the numerical and WKB results. We investigate the prefactors in Sec. \ref{sec:prefactor}.

\subsubsection{Non-analytic behavior of the entropic barrier near the Hopf bifurcation}
\label{sec:bifurcation}

\begin{figure}
\includegraphics[width=1.81 in,clip=]{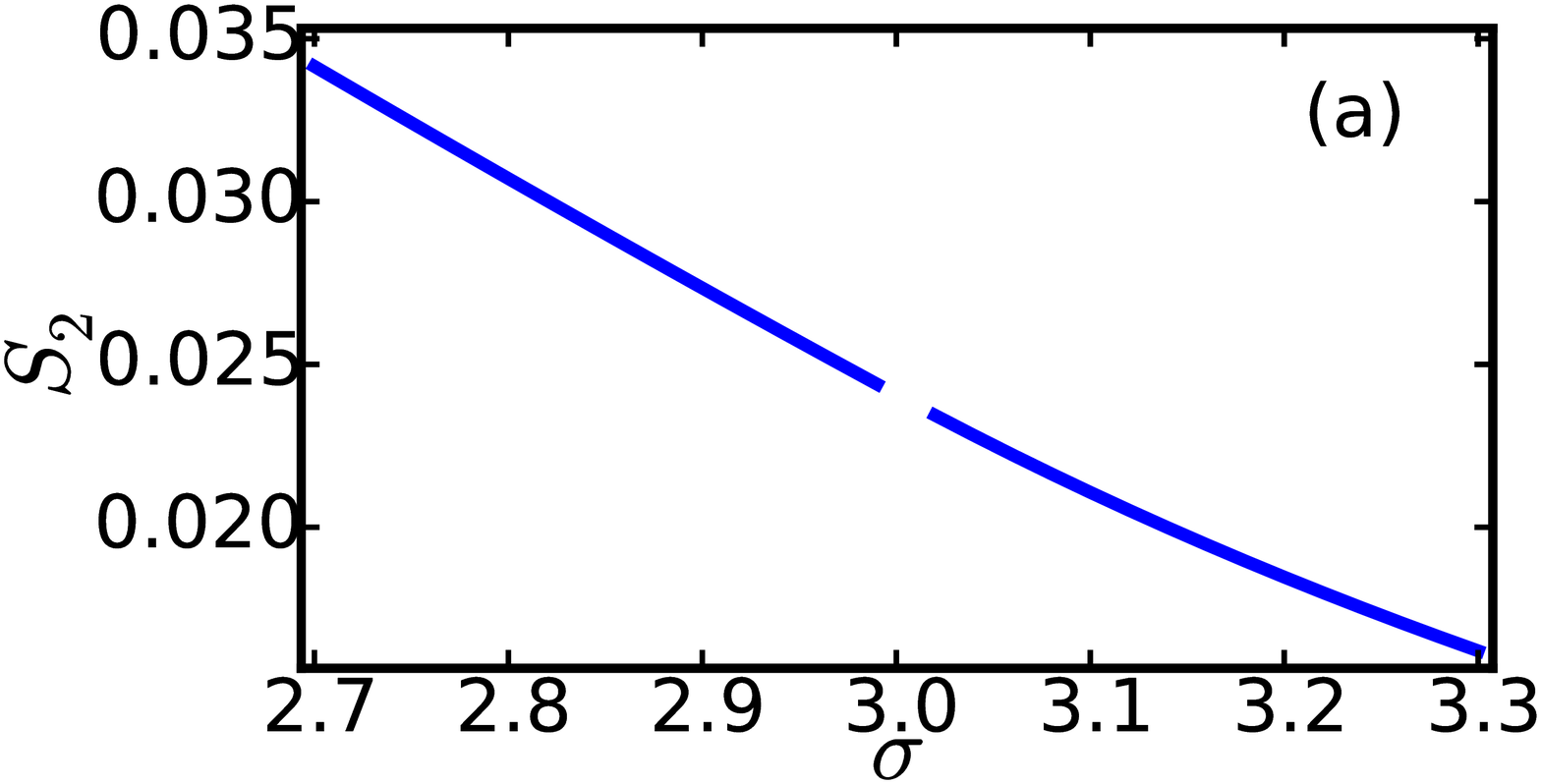}
\includegraphics[width=1.55 in,clip=]{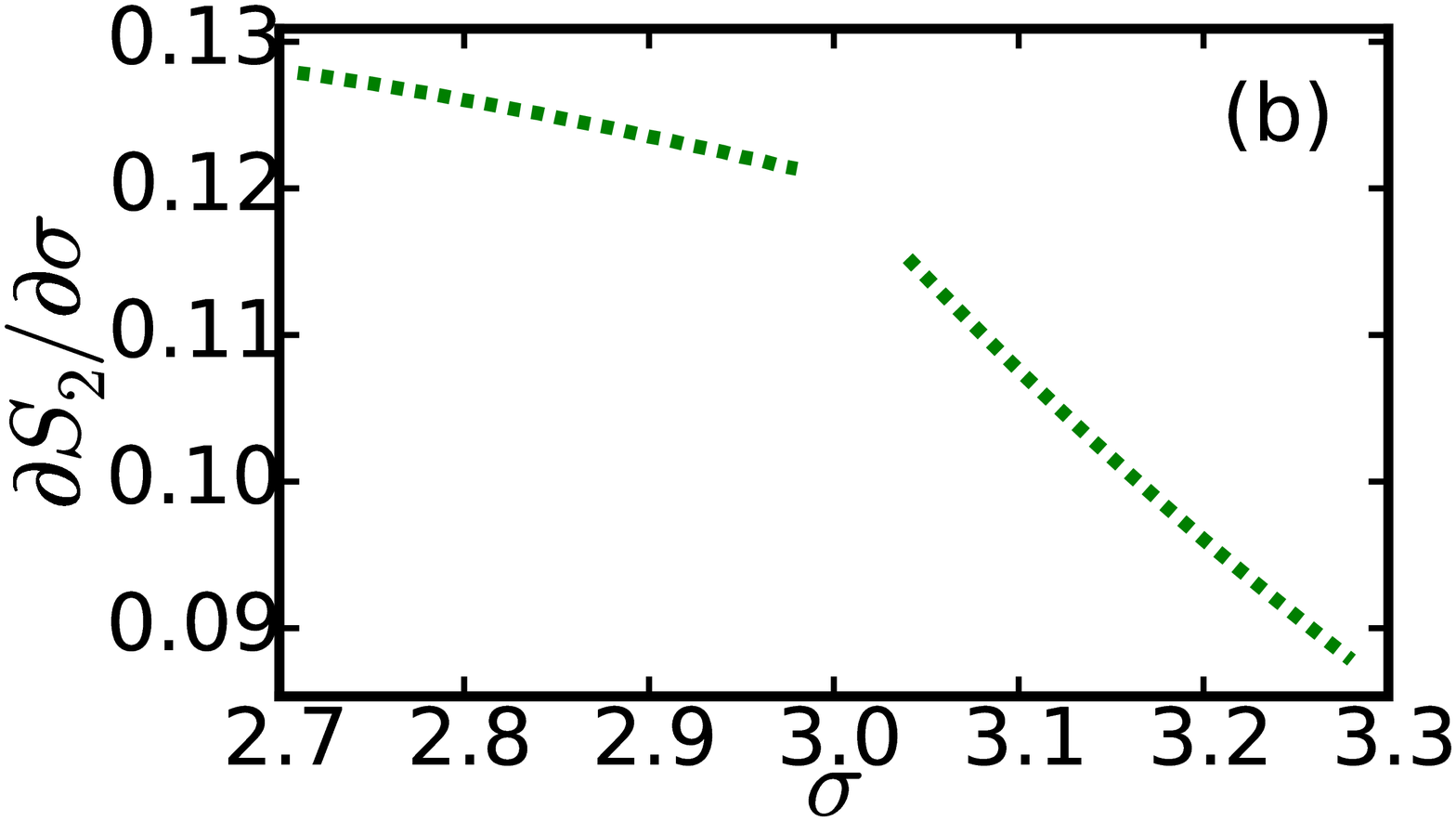}
\includegraphics[width=1.64 in,clip=]{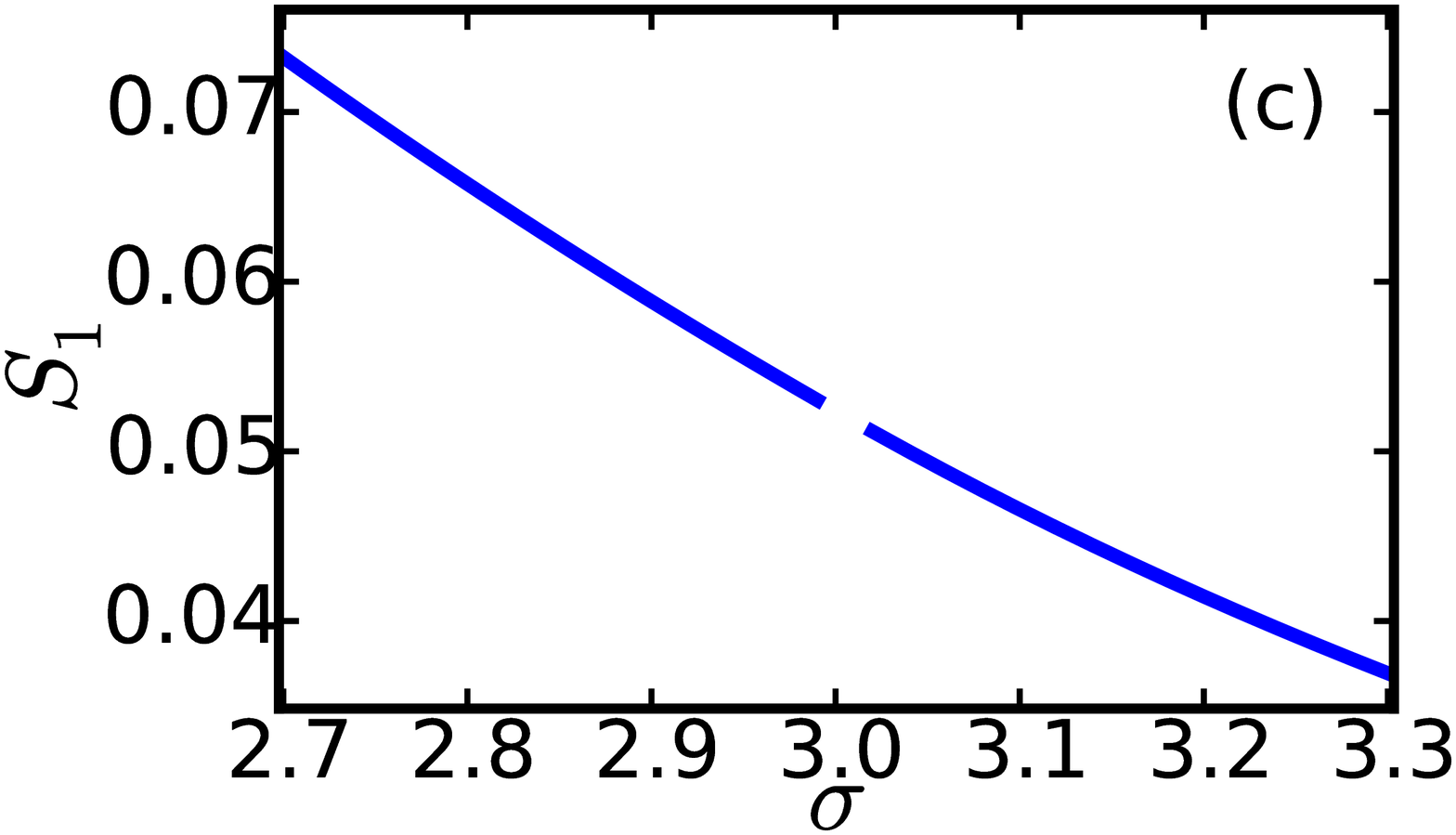}
\includegraphics[width=1.72 in,clip=]{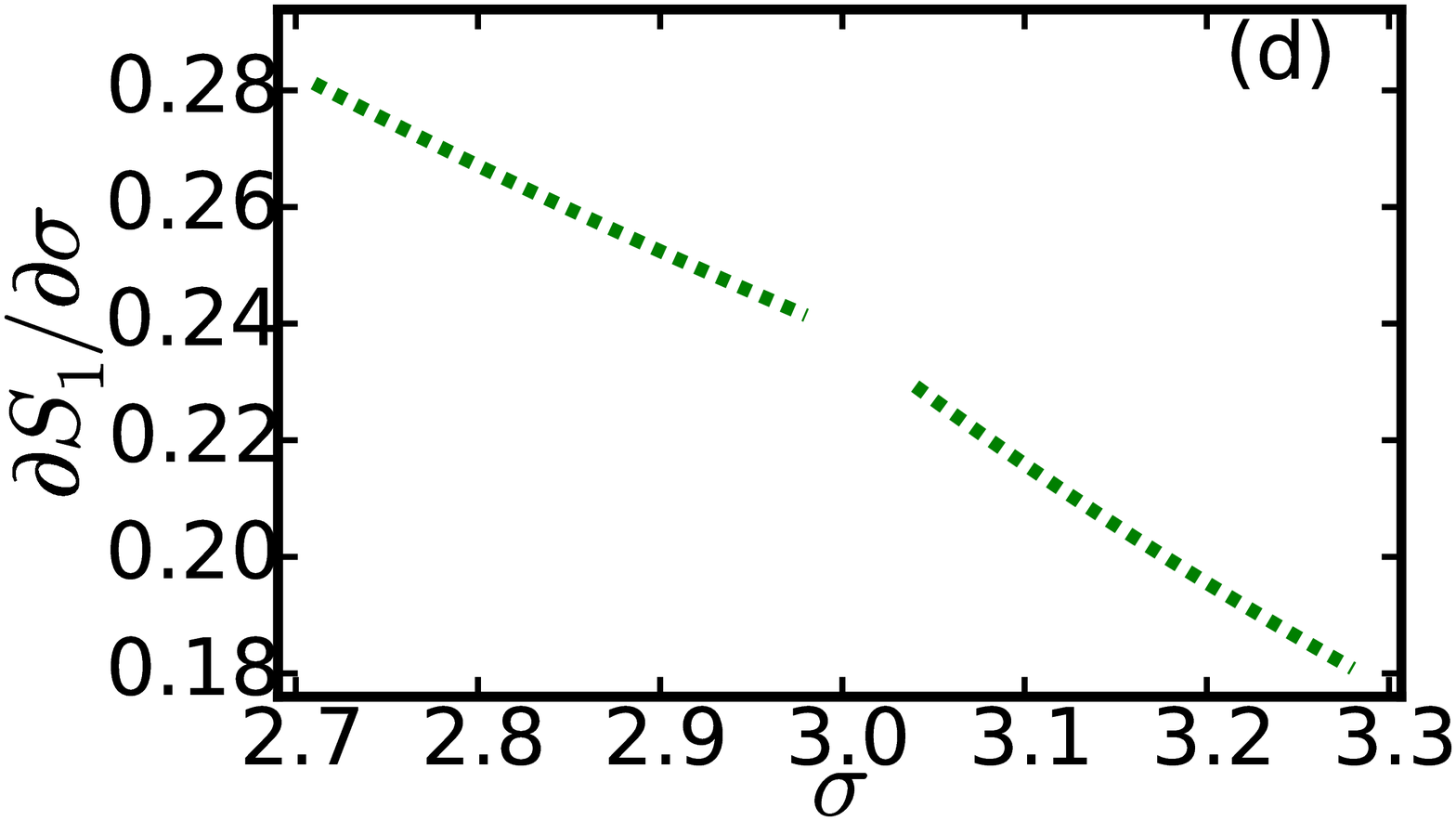}
\caption{(Color online) Jumps in the second derivatives of the entropic barriers to extinction with respect to the bifuraction parameter $\sigma$ at the Hopf bifurcation. Solid lines: the actions $S_2$ (a) and $S_1$ (b) along the instantons. Dashed lines: the first derivatives $\partial S_{2}/\partial\sigma$ (a) and $\partial S_{1}/\partial\sigma$ (d).  The actions and their first derivatives are continuous, while the second derivatives are discontinuous at the Hopf bifurcation. The parameters are $a=1, \tau=0.5$. The Hopf bifurcation occurs at $\sigma^*=3$, see Eq.~\ref{eq_sigma_star}.}
 \label{fig_bif}
\end{figure}

Figure \ \ref{fig_bif} presents our numerical results for the actions $S_{1,2}$ calculated along the instantons leading to $F_1$ and $F_2$, respectively, for the same $a=1$ and $\tau=0.5$ and different values of $\sigma$ close to the Hopf bifurcation, $\sigma=\sigma^{*}$. For $\sigma < \sigma^*$ the fixed point $M_3$ is stable, and the actions were calculated on the instantons which start at $M_3$. An immediate observation is that  $S_{1}>S_{2}$ for all $\delta$, so that that the effective transition rate $\mathcal{R}_{2}$ is \emph{exponentially} greater than $\mathcal{R}_{1}$, as observed previously for extinction from a fixed point \cite{GM2012}. The (numerically evaluated) first derivative of the entropic barriers with respect to $\sigma$ are also shown as the function of $\sigma$ (figures b and d). Interestingly, it exhibits a corner singularity at $\sigma=\sigma^*$, indicating a jump in the second derivative (so the phenomenon can be classified as a dynamic second-order phase transition).

Let us now consider a simple (and well known) toy model of a ``noisy" Hopf bifurcation that displays the same
phenomenon but can be solved analytically. The model is defined by two Langevin equations in polar
coordinates:
\begin{equation}
\label{eq_langevin_toy_model}
\dot{r}=\alpha r-\beta r^{3}+\gamma r^{5}+\epsilon\xi\left(t\right)\qquad\dot{\theta}=1,
\end{equation}
where $\xi\left(t\right)$ is a zero-mean Gaussian noise, delta-correlated
in time.
As the equations for $r$ and $\theta$ are decoupled,
the problem is effectively 1-dimensional.
We assume that $\beta$ and $\gamma$ are positive, and examine the Hopf bifurcation
which occurs when $\alpha$ changes sign. Close to the bifurcation $\left|\alpha\right|\ll1$.
We consider the weak-noise limit, $\epsilon\rightarrow0$ and assume that the noise is smaller than the rest of parameters, including  $\left|\alpha\right|$.

\begin{figure}
\includegraphics[width=3.2 in, clip=]{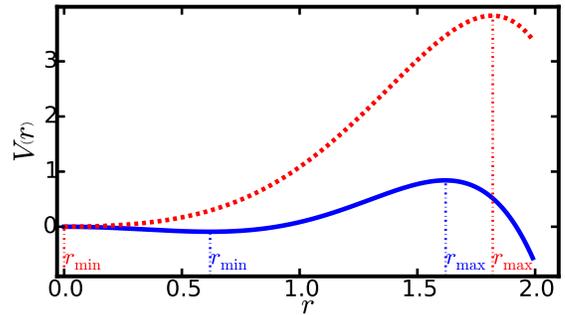}
\caption{(Color online) The deterministic potential (\ref{eq_toy_model_potential}) of the toy model. The parameters are $\beta=3$, $\gamma = 1$, and $\alpha=1$ (solid) and $\alpha=-1$ (dashed). $r_{\min}$ and $r_{\max}$ are marked on the figure. The Hopf bifurcation is at $\alpha = 0$, where $r_{\min}$ changes from zero to a positive value. Close to the bifurcation $\left|\alpha\right|\ll1$. However, for the clarity of the figure we used values of $\alpha$ which are not small.} \label{fig_toy model}
\end{figure}

The equation for $r$ can be written in the form $\dot{r}=-V'\left(r\right)+\epsilon\xi\left(t\right)$,
where we have defined the deterministic potential
\begin{equation}
\label{eq_toy_model_potential}
V\left(r\right)=-\frac{\alpha}{2}r^{2}+\frac{\beta}{4}r^{4}-\frac{\gamma}{6}r^{6},
\end{equation}
see Fig. \ref{fig_toy model}.

In the toy model, the mean escape time (MET) from the metastable state near the origin $r_{\min}$ %to the stable state $r\rightarrow\infty$
is given by the Kramers' formula \citep{Gardiner1985}. To leading order, $\ln\left(\text{MET}\right)\simeq 2\Delta V/\epsilon^{2}$,
where $\Delta V=V\left(r_{\max}\right)-V\left(r_{\min}\right)$ is the
height of the potential barrier, $r_{\max}$ is the global maximum
of $V$, and $r_{\min}$ is the coordinate of the metastable state near the origin, see Fig. \ref{fig_toy model}.  The quantities $\Delta V$ and $1/\epsilon^{2}$ are analogous
to the action $S$ and the characteristic population size $N$, respectively, in the stochastic RMA model.

For $\alpha<0$, the origin is a stable fixed point of the deterministic
dynamics, and a local minimum of $V$, so $r_{\min}=0$, and $V\left(r_{\min}\right)=0$.
For $\alpha>0$, the origin is unstable, and a limit
cycle appears around it, of radius $r_{\min}\simeq\sqrt{\alpha/\beta}$. Now the local minimum of the potential is $V\left(r_{\min}\right)\simeq-\alpha^{2}/\left(4\beta\right)$, where we have assumed that $r_{\min} \ll 1$, which holds near the bifurcation.

If we now consider $\Delta V$ as a function of $\alpha$, we can
easily see that it is non-analytic at $\alpha=0$, because of the non-analytic behavior of
$V\left(r_{\min}\right)$. From the expressions for $V\left(r_{\min}\right)$
below and above the bifurcation, we find that the second derivative $\partial^{2}\Delta V/\partial\alpha^{2}$
jumps by $1/\left(2\beta\right)$.

This prediction from the toy model is quite general, and only depends on the
behavior of the potential near the origin. In particular, the value (and even the sign) of $\gamma$ in Eq.~\ref{eq_toy_model_potential} does not affect the results in any way -- what only matters is that higher order terms in $r$ guarantee
the existence of a finite $r_{\max}$.

The prediction of non-analytic behavior at the Hopf bifurcation can be extended to the stochastic RMA model. Near the bifurcation, and in the vicinity of $M_3$ and/or the limit-cycle around it, the Fokker-Planck approximation of the master equation (\ref{MasterEquation}) is applicable and can be obtained by the van Kampen system-size expansion \cite{Gardiner1985}. After rescaling the coordinates, a radial Langevin equation can be deduced, and it is of the same form as Eq.~(\ref{eq_langevin_toy_model}).  However, the amount by which  $\partial^{2}S_i/\partial\sigma^{2}$ jumps  in the stochastic RMA model is in disagreement with the prediction of the toy model. To begin with, the jumps of the second derivatives for $S_1$ and $S_2$ are in general not equal to each other, as clearly seen in Fig. \ref{fig_bif}. The reason for the disagreement is that, in a non-equilibrium system like the
stochastic RMA model, the actions $S_1$ and $S_2$ are nonlocal, and the jumps in their second derivatives with respect to the $\sigma$ are affected by the variation along the entire instantons. Therefore, the toy model misses one crucial feature of the stochastic RMA model: its non-equilibrium character.

\subsubsection{Prefactors of the extinction rates}
\label{sec:prefactor}

\begin{figure}
\includegraphics[width=3.3 in,clip=]{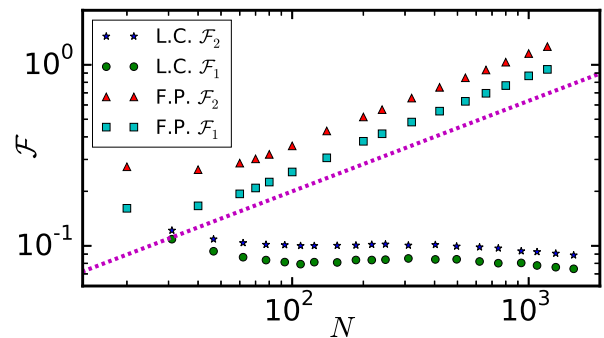}
\caption{The $N$-dependence of the numerically evaluated prefactors $\mathcal{F}_{i}\left(a,\sigma,\tau,N\right)$, ($i=1,2$), for fixed $a=1$, $\tau=0.5$, and $\sigma=3.1$ (limit cycle -- L.C.) and $\sigma=2$ (fixed point -- F.P.). For $N\gg1$, there is an apparent power-law dependence on $N$, whose exponent changes at the bifurcation. For a fixed point, the prefactor scales as $\mathcal{F}\propto N^{1/2}$, while for a limit cycle it appears to be independent of $N$. The dashed line has a slope of $1/2$ and is meant to guide the eye.}
\label{fig_prefactor}
\end{figure}

We now study the
prefactors $\mathcal{F}_1$ and $\mathcal{F}_2$, which are the subleading corrections to the extinction rates $\mathcal{R}_{i}=\mathcal{F}_{i}e^{-NS_{i}}$, $i=1,2$ at $N\gg 1$ . Our numerical results for the dependence of the prefactors on $N$ (for fixed $a, \sigma,$ and $\tau$) are shown in Fig.\ \ref{fig_prefactor}. One can see an apparent power-law dependence, whose exponent changes at the Hopf bifurcation $\sigma = \sigma^*$. We observed
the same behavior for other sets of $a, \sigma, \tau$ as well (not shown). When the fixed point $M_3$ is attracting, the prefactors appear to behave as $\mathcal{F} \propto N^{1/2}$. For a stable limit cycle, they appear to be independent of $N$.

How can we interpret these numerical results? For multi-population systems the prefactors cannot, in general, be found analytically. It is natural to assume, however, that the prefactors depend on $N\gg 1$ as a power-law,  $\mathcal{F} \propto N^{\mu}$. Indeed, apart from our numerical results for the RMA model, power-law behaviors of the prefactors of the extinction rates have been observed in all cases where the prefactors could be calculated analytically:
for single populations \cite{Assaf2010metastable,AM2006,Kessler2007,Assaf2010}  and for two-population systems which possess a time-scale separation \cite{KMS2010}.

We shall now propose a simple argument which explains why the exponent of the power-law $\mu$ changes by $1/2$ at the Hopf bifurcation, as observed in Fig.\ \ref{fig_prefactor}. The argument is based on the normalization of the quasi-stationary distribution $\pi_{mn}$, which is strongly affected by the Hopf bifurcation.

When $M_3$ is a stable fixed point, the quasi-stationary distribution $\pi_{mn}$ in the vicinity of $M_3$ is a two-dimensional Gaussian peaked at $M_3$, with standard deviations of order $\sqrt{N}$ in either direction. This part of the distribution gives the main contribution to the normalization. The value of the quasi-stationary distribution function at its peak is therefore of order $\pi_{m^{*}n^{*}}\sim1/N$, where $\left(m^{*},n^{*}\right)=\left(Nx^{*},Ny^{*}\right)$ is the fixed point $M_3$.

In the case of a limit cycle, the distribution $\pi_{mn}$ is well approximated by a Gaussian ``ridge" around the limit cycle, whose width is of order $\sqrt{N}$ \citep{Dykman1993}. Since the length of the limit cycle increases linearly with $N$, the value of the quasi-stationary distribution function on the limit cycle is of order $\pi\left(m_{lc},n_{lc}\right)\sim N^{-3/2}$. This will in turn add a factor of order $N^{-1/2}$ to the whole distribution $\pi_{mn}$, compared to the fixed point case.  We therefore expect the prefactor's power-law dependence on $N$ to change at the bifurcation, and the exponent $\mu$ to drop by $1/2$ at the Hopf bifurcation, as indeed observed in Fig.\ \ref{fig_prefactor}.

\section{Conclusions}
\label{sec:discussion}
We studied population extinction from a limit cycle due to intrinsic noise in the Rosenzweig-MacArthur model. In the leading-order WKB approximation, the calculation of the extinction rates and the most probable extinction paths boils down to finding a new type of instantons of an effective Hamiltonian system. We developed a numerical method of computing the instantons. The method is based on Floquet theory and involves numerical determination of the unstable direction by which the instantons exit the limit cycle.

We showed numerically that the entropic barriers to extinction $S_1$ and $S_2$ exhibit a non-analytic behavior -- a second-order dynamic phase transition -- at the Hopf bifurcation. We leave, as an open problem, the challenge of analytically calculating the jump in the second derivative. We believe that there are two contributions to this jump: one is a local contribution from the immediate vicinity of the fixed point $M_3$ that can be calculated analytically, as we have shown by using a simple toy-model. The second contribution comes from the variation of the action due to the global change of the shape of the instanton. This one is hard to calculate analytically.

We also obtained numerical results for the prefactors -- the subleading corrections to the WKB extinction rates. The prefactors show a power-law dependence on the population size scale $N$.  We found that the power law changes at the Hopf bifurcation, and suggested an explanation of this change in terms of a simple normalization argument.

\section*{ACKNOWLEDGMENTS}

We are very grateful to Dr. Alberto d'Onofrio who attracted our interest in the population extinction from a limit cycle. This work was supported by the
US-Israel Binational Science Foundation (Grant No.
2008075).

\end{document}